\documentclass[aps,prd,superscriptaddress,amsfonts,amssymb,amsmath,nofootinbib,showpacs,showkeys]{revtex4}
\usepackage{ulem,color,wrapfig,graphicx}

\newcommand{\rem}[1]{}
\usepackage{amsfonts,amssymb,amsmath,amsthm,mathrsfs}
\usepackage{url}
\usepackage{graphicx}
\newtheorem{main}{ Theorem}
\newtheorem{thrm}{Theorem}[section]
\newtheorem{lem}[thrm]{Lemma}
\newtheorem{prop}[thrm]{Proposition}

\newtheorem{remark}[thrm]{Remark}
\theoremstyle{definition}

\def\XXint#1#2#3{{\setbox0=\hbox{$#1{#2#3}{\int}$}
     \vcenter{\hbox{$#2#3$}}\kern-.5\wd0}}

\begin{document}

\author{Salvatore Capozziello}
\email{capozziello@na.infn.it}
\address{Dipartimento di Fisica ``E. Pancini", Universit\`a di Napoli {\it   Federico II}, Napoli,\\ and
INFN Sez. di Napoli, Compl. Univ. di Monte S. Angelo, Edificio G, Via Cinthia, I-80126, Napoli, Italy,}
\address{Laboratory for Theoretical Cosmology, Tomsk State University of Control Systems and Radioelectronics (TUSUR), 
\\634050 Tomsk (Russia).}

\author{Carlo Alberto Mantica}
\email{carlo.mantica@mi.infn.it}
\address{I.I.S. Lagrange, Via L. Modignani 65, 
I-20161, Milano, Italy \\
and INFN sez. di Milano,
Via Celoria 16, I-20133 Milano, Italy}

\author{Luca Guido Molinari}
\email{luca.molinari@unimi.it}
\address{Dipartimento di Fisica ``A. Pontremoli'',
Universit\`a degli Studi di Milano\\ and INFN sez. di Milano,
Via Celoria 16, I-20133 Milano, Italy.}
\email{luca.molinari@unimi.it}


\title{General properties of $\mathbf {f(R)}$ gravity  vacuum solutions}

\begin{abstract}
General properties of vacuum solutions of $f(R)$ gravity are obtained by the condition that the divergence of the Weyl tensor is zero and $f''\neq 0$. Specifically, a theorem states that  the gradient of the curvature scalar, $\nabla R$, is an eigenvector of the Ricci tensor and, if it  is time-like, the space-time is a Generalized Friedman-Robertson-Walker metric; in dimension four, it is Friedman-Robertson-Walker. 
\end{abstract}

\date{\today}
\maketitle

\section{Introduction}
The so called $f(R)$ gravity is a natural extension of Einstein's gravity where the Hilbert-Einstein action of gravitational field, linear in the Ricci scalar $R$, is substituted with a generic function $f(R)$.  The issues for this generalization mainly come from inflationary cosmology \cite{Starobinsky},  late-time acceleration \cite{curvature}  and the possibility to unify late and early cosmic history \cite{Odintsov,Oikonomou}.\\
Furthermore, it is the subject of a vast research as a
potential alternative to the so far undetected exotic fields that should account for  dark matter and 
dark energy. This alternative is geometric: the further degrees of freedom of $f(R)$ gravity   may produce  observable effects   at different astrophysical and cosmological scales that should be,  otherwise, ascribed to exotic forms of matter.
There are other motivations like  quantum perturbative corrections  on curved spacetimes and the natural question about the  consequences of a straightforward generalization of  the  Hilbert-Einstein action to consider $f(R)$ gravity as the first logical step
\cite{CapFar11,CapDeL11}. 

Starting from a general $f(R)$ gravity action, the field equations  can be written as  
\begin{equation}
G[f]_{kl} = \kappa T_{kl}\,,
\end{equation}
where $G[f]$ replaces the Einstein 
tensor:
\begin{align}
G[f]_{kl} = f'(R)R_{kl} &-f'''(R)\nabla_k R\nabla_l R - f'' (R) \nabla_k\nabla_l R \label{Gf}\\
 &+ g_{kl}[f'''(R)\nabla_jR\nabla^j R + f''(R) \nabla^2 R-\tfrac{1}{2}f(R)]\,.\nonumber
 \end{align} 
It becomes the standard expression, $G_{kl}=R_{kl}-\tfrac{1}{2}Rg_{kl}$, for $f(R)=R$. $T_{kl}$ is the stress-energy tensor for  standard matter and $\kappa$ is the gravitational coupling. 
A main requirement is that the contracted Bianchi identity
$\nabla^k G_{kl}=0$ continues to hold in $f(R)$ gravity, that is:
\begin{align*} 
\nabla^k G[f]_{kl}= f''(R)[ R_{kl}\nabla^k R - \nabla^2 \nabla_l R  + \nabla_l \nabla^2 R ] =0\,. 
\end{align*}
Therefore $f(R)$ theories are compatible with the physical requirement of energy-momentum conservation.\\
In Ref.\cite{CMM2018}  we obtained a sufficient condition for $G[f]_{kl}$ to have the `perfect-fluid' form
\begin{equation}
G[f]_{kl} = g_1(R) g_{kl} + g_2(R) u_k  u_l \,,
\end{equation}
for any smooth $f(R)$ model, where $u_k$ is some time-like unit vector field. The requirement that each tensor
in the expression \eqref{Gf} of $G[f]_{kl}$ have the perfect fluid form, namely the Ricci tensor $R_{kl}$,  $\nabla_k R \nabla_l R$ 
and $\nabla_k\nabla_l R$, can be stated as follows:

C1) $\nabla_k u_j = \varphi (u_j u_k + g_{jk}) $ with $\nabla_k \varphi =-\dot\varphi u_k$,

C2) $\nabla_m C_{jkl}{}^m=0$, where $C_{jkl}{}^m$ is the Weyl tensor.\\
Condition C1 characterizes the space-time as a generalized Friedman-Robertson-Walker (GFRW) space-time.  
The additional condition 2 implies that $u_k\propto \nabla_k R$. Then $\nabla_k R$ is eigenvector of the Ricci tensor.\\ 
In  $n=4$ dimensions, C1 and C2 imply that the space-time is Friedman-Robertson-Walker \cite{MMJMP} Prop. 4.1. 
The unicity of the vector $u_k$ realizing a GFRW space-time is
discussed in \cite{MTM}.

The perfect-fluid structure of $G[f]_{kl}$ and  $T_{kl}$,  required in large-scale cosmology, put $f(R)$ geometric extension  of General Relativity on the same footing as exotic modifications of standard matter. In other words, the question is whether cosmic phenomenology can be addressed by requiring exotic forms of matter/energy beyond the Standard Model of Particles, or gravity is not scale invariant and modifications are required at galactic scales and beyond \cite{Borka1,Borka2}.

Hereafter, we investigate  vacuum solutions ($T_{kl}=0$) in space-times of dimension $n$ in order to derive some general properties of $f(R)$ gravity.\\
Vacuum solutions have been studied based on special forms of $f(R)$ or assumptions about symmetries of the metric \cite{Stabile}. Perturbations of models like  $f(R)=R^{1+\delta}$ in spherical symmetry are studied in \cite{Cli06,Far09}. The non-linear equations for $f'$ for
spherical-symmetric vacuum solutions are studied in \cite{Car09}, and solved for some cases.
Constant curvature solutions, i.e. $f'(R)R-2f(R)=0$, were found with cylindrical symmetry \cite{Aza08,Mom09},
plane-symmetry \cite{Sharif10}, and local rotational symmetry \cite{Amir14}. Spherically symmetric solutions 
in connection to black-holes and the rotation curve of galaxies are studied in \cite{Calza18}.

In this work we want to characterize the vacuum solutions of any smooth $f(R)$ theory with null divergence of the  Weyl tensor. The results can be summarized  in the following theorem:
\begin{main}
For any smooth function $f(R)$ with $f''\neq 0$, the vacuum solution with null divergence of the Weyl tensor ($\nabla_m C_{jkl}{}^m=0$), has the properties:\\
1) $\nabla_k R$ is an eigenvector of the Ricci tensor: $R_j{}^k \nabla_k R=\xi \nabla_j R$.\\ 
2) If $\nabla_k R=\alpha u_k$ with $u^ku_k=-1$,  the Ricci tensor has the perfect fluid (quasi-Einstein) form:
\begin{align}
 R_{kl} = \frac{R-n\xi}{n-1} u_k u_l +\frac{R-\xi}{n-1} g_{kl} \,. \label{main1}
\end{align}
3) The vector $u_k$ is vorticity-free, acceleration-free and shear-free, and satisfies the relation:
\begin{align}
\nabla_j u_k =\varphi (u_j u_k+g_{jk})  \label{nabla}\,,
\end{align}
where $\nabla_k\varphi =-\dot\varphi u_k $. 
\end{main}
\noindent
The property \eqref{nabla} characterizes the Lorentzian space-time as a
generalised Robertson-Walker space-time. It is usually described as a
time-warped space-time, i.e. there is a coordinate frame where the metric tensor is:
\begin{align}
ds^2 = -dt^2 + a^2(t) g^*_{\mu\nu} (x) dx^\mu dx^\nu\,,
\end{align}
and $g^*$ is a Riemannian metric \cite{review, Chen}. For such space-times we proved
the special property $C_{jkl}{}^m u_m =0$ iff $\nabla_m C_{jkl}{}^m=0$ \cite{MMJMP}.
In $n=4$ it implies that the Weyl tensor is zero if the divergence vanishes, i.e.
the space-time is Robertson-Walker.\\

The proof of the above Theorem is given in the next sections. In Sect.2, we discuss  $\nabla_k R$ which is an eigenvector of the Ricci tensor. In Sect.3 the related form of Ricci tensor is obtained. Sect.4 is devoted to the vector 
$u_k$ which is vorticity and acceleration-free; its shear is
proportional to the electric component of the Weyl tensor $\sigma_{kl}\propto C_{jklm}u^ju^m$. In 
Sect.5 an equation for the time evolution of the shear is obtained.
A vorticity and acceleration free velocity field restricts the metrics to the form 
$ds^2=-dt^2+a(t,x)_{\mu\nu}dx^\mu dx^\nu$. By the equations in Appendix we show, in Sect.6, that the 
special form of the shear implies that it is zero.
The consequences are discussed in the conclusions (Sect.7). All results are then simplified because the Ricci tensor is quasi-Einstein, the Ricci tensor of the space-submanifold is Einstein, the space-time is a GFRW. In Appendix are reported all the  quantities used for the derivations developed in the paper. We indicate with  a dot the directional derivative $u^k\nabla_k$  and with a prime the   derivative with respect to  $R$.

\section{$\mathbf {\nabla_k R}$ is an eigenvector of the Ricci tensor}
 Let us rewrite the field equations $G[f]_{kl}=0$, with $G[f]$ given in Eq.\eqref{Gf}, as:
 \begin{gather}
 R_{kl} = A \nabla_k R\nabla_l R + B \nabla_k\nabla_l R- C g_{kl} \,,\label{Ricci}\\
 \mbox{where}\;\;\;\;A=\frac{f'''}{f'}, \; B=\frac{f''}{f'}, \; C= A\nabla_jR\nabla^j R + B \nabla^2 R-\frac{f}{2f'}\,. \nonumber
 \end{gather}
 The trace is: 
 \begin{align}
 R = -(n-1)[A \nabla_k R\nabla^k R + B \nabla^2 R] + n \frac{f}{2f'} \label{curvature}\,.
 \end{align}
 Then
 \begin{align}
  (n-1)C = -R +\frac{f}{2f'} \label{C}\,.
 \end{align}
 \noindent
 {\bf Proof}: let us evaluate
  \begin{align*}
 \nabla_j R_{kl} = (A' \nabla_j R)\nabla_k R\nabla_l R + A(\nabla_j\nabla_k R)\nabla_l R 
 +A\nabla_k R (\nabla_j\nabla_lR)\\
 + (B' \nabla_j R)\nabla_k\nabla_l R + B\nabla_j\nabla_k\nabla_l R -C' g_{kl} \nabla_j R\,.
 \end{align*}
 As said, the prime indicates the derivative in $R$ and we antisymmetrize in two indices:
  \begin{align*}
 \nabla_j R_{kl} -\nabla_k R_{jl} =& (A-B')[ (\nabla_k R) \nabla_j\nabla_lR- (\nabla_j R) \nabla_k\nabla_l R]
 \\
 & +B R_{jkl}{}^m \nabla_m R -C' (g_{kl} \nabla_j R - g_{jl}\nabla_k R)\,.
 \end{align*}
 The divergence of the Weyl tensor is
 \begin{align*}
 \nabla_m C_{jkl}{}^m = -\frac{n-3}{n-2}\left[ \nabla_j R_{kl}-\nabla_k R_{jl}-
\frac{g_{kl}\nabla_j R -g_{jl}\nabla_k R}{2(n-1)}\right ]\,.
 \end{align*}
 If $\nabla_m C_{jkl}{}^m =0$ then:
\begin{align*}
\frac{g_{kl}\nabla_j R -g_{jl}\nabla_k R}{2(n-1)} =& (A-B')[ (\nabla_k R) \nabla_j\nabla_lR- (\nabla_j R) \nabla_k\nabla_l R]\\
 & +B R_{jkl}{}^m \nabla_m R -C' (g_{kl} \nabla_j R - g_{jl}\nabla_k R)\,.
 \end{align*}
Note that $A-B'=B^2$ and $C'=-\frac{1}{2(n-1)}-\frac{1}{2(n-1)}\frac{f}{f'}B$
 \begin{align*}
-B\frac{g_{kl}\nabla_j R -g_{jl}\nabla_k R}{2(n-1)} \frac{f }{f'}  = B^2[ (\nabla_k R) \nabla_j\nabla_lR- (\nabla_j R) \nabla_k\nabla_l R]  +B R_{jkl}{}^m \nabla_m R\,.
  \end{align*}
 We can factor out $B\neq 0$. 
 The term $B\nabla_j\nabla_lR $ is obtained from the expression \eqref{Ricci} of the Ricci tensor. In this
 substitution,  terms with $A$ cancel out:
 \begin{align*}
-\frac{g_{kl}\nabla_j R -g_{jl}\nabla_k R}{2(n-1)} \frac{f }{f'}  =  (\nabla_k R) (R_{jl} +Cg_{jl}) - (\nabla_j R)(R_{kl} +Cg_{kl}) +R_{jkl}{}^m \nabla_m R\,. 
 \end{align*}
A simplification occurs with Eq.\eqref{C}:
\begin{align}
0 =  (\nabla_k R) (R_{jl} -\frac{R}{n-1}g_{jl}) - (\nabla_j R)(R_{kl} -\frac{R}{n-1}g_{kl}) +R_{jkl}{}^m \nabla_m R \,.
\label{Riem}
 \end{align}
The contraction with $\nabla^l R$ cancels out terms, leaving
$(\nabla_k R) R_{jl} \nabla^l R  = (\nabla_j R) R_{kl}\nabla^l R$. The equation is solved by: 
\begin{align}
R_{jl}\nabla^l R = \xi \nabla_j R\,,
\end{align}
for some eigenvalue $\xi$. This completes the proof of point 1 of the Theorem 1. \qquad $\square $

\section{The Ricci tensor}
Let us  obtain the structure of the Ricci tensor with the approach  used in \cite{MMJMP}. The result will be 
simplified after showing that the shear of $u_k$ is zero. \\
In the following, we refer to a time-like unit vector: $\nabla_k R = \alpha u_k $,  where $u^k u_k =-1$. 
$u^k$ is an eigenvector of the Ricci tensor, $R_{kj}u^j =\xi u_k$. 
 Eq.\eqref{Riem} is
\begin{align}
R_{jklm} u^m  =  -u_k (R_{jl} -\frac{R}{n-1}g_{jl}) + u_j(R_{kl} -\frac{R}{n-1}g_{kl})\,.  \label{Riem2}
 \end{align}
Contracting  with $u^j$, gives:
\begin{align}
R_{jkl}{}_m u^j u^m  =  - \xi u_k u_l  - R_{kl} + \frac{R}{n-1}(g_{kl}+u_ku_l)\,.  \label{Riem3}
\end{align}
The Weyl tensor 
\begin{align*}
C_{jklm} = R_{jklm} + \frac{ g_{jm} R_{kl} - g_{km} R_{jl} + g_{kl}R_{jm} -
g_{jl} R_{km}}{n-2} - R \frac{ g_{jm}g_{kl}-
g_{km}g_{jl} }{(n-1)(n-2)} \,,
\end{align*}
is contracted with $u^ju^m$ to obtain the Ricci tensor, and Eq.\eqref{Riem3} is used. It is
\begin{align*}
(n-2)C_{jklm} u^ju^m=& (n-2)R_{jklm}u^ju^m 
-R_{kl} - \xi (2u_k u_l +g_{kl})  + R \frac{ g_{kl}+u_ku_l }{n-1} \\
=& (R - n\xi) u_k u_l  -(n-1) R_{kl} + (R-\xi )g_{kl} 
\end{align*}
The resulting Ricci tensor has a quasi-Einstein term and a Weyl term: 
\begin{align}
R_{kl} = \frac{R - n\xi}{n-1} u_k u_l  + \frac{R-\xi}{n-1} g_{kl}  -\frac{n-2}{n-1}C_{jklm} u^ju^m \label{Ricci2}
\end{align}
The Weyl term will be shown to be zero.

\section{The vector field $\mbox{\large $\mathbf {u_k}$} $ }
Let us obtain now the properties of the vector field $u_k$  ($\nabla_k R=\alpha u_k$).\\
We  rewrite the Ricci tensor \eqref{Ricci} in terms of $u_k$:
\begin{align*}
R_{kl} = A \alpha^2 u_k u_l  + B (\nabla_k \alpha) u_l + B\alpha \nabla_k u_l  - C g_{kl} \,.
\end{align*}
The contraction with $u^l$ gives: $\xi u_k =-A\alpha^2 u_k -B(\nabla_k \alpha) -Cu_k$. Then $\nabla_k\alpha $
is proportional to $u_k$:
\begin{align}
\nabla_k \alpha = -\dot\alpha u_k   \label{dotalpha}
\end{align}
and the Ricci tensor is:
\begin{align}
R_{kl} = (A \alpha^2  - B\dot\alpha )u_k u_l + B\alpha (\nabla_k u_l)  - C g_{kl} \,.
\end{align}
\begin{lem}
The vector field $u_k$ is vorticity-free and acceleration-free.
\begin{proof}
Eq.\eqref{dotalpha} and the identity $\nabla_k\nabla_j R=\nabla_j \nabla_k R$, i.e. $\nabla_k (\alpha u_j) = \nabla_j (\alpha u_k)$,  give
\begin{align}
\nabla_j u_k - \nabla_k u_j =0\,.
\end{align}
Contraction with $u^j$ gives zero acceleration: $u^j\nabla_j u_k = 0$ because $u^j u_j=-1$. 
\end{proof}
\end{lem}
This Lemma has the consequence that the gradient of $u_k$ has the structure  
\begin{align}
\nabla_j u_k =\varphi (u_j u_k+g_{jk}) +\sigma_{jk} \,,
\end{align}
with $\varphi =  \frac{\nabla_i u^i}{n-1}$, and shear tensor $\sigma_{jk}$ 
(traceless, symmetric and $\sigma_{jk}u^k=0$). 
The Ricci tensor becomes:
\begin{align}
R_{kl} = (A \alpha^2  - B\dot\alpha +B\alpha\varphi )u_k u_l + (B\alpha \varphi  - C) g_{kl}  + B\alpha \sigma_{kl} \,.
\end{align}
Comparison with the expression \eqref{Ricci2} gives 
$$B\alpha \sigma_{kl} = -\frac{n-2}{n-1}C_{jklm} u^ju^m \,,$$
and the relations
\begin{align*}
&R-n\xi =(n-1)(A \alpha^2  - B\dot\alpha +B\alpha\varphi )\\
&R-\xi = (n-1)(B\alpha \varphi  - C)\,.
\end{align*}
The second one simplifies with Eq.\eqref{C}, and it is used in the first one:
\begin{align}
&(n-1)B\alpha \varphi = \frac{f}{2f'}-\xi     \label{F1}    \\
&(n-1)(A\alpha^2 -B\dot \alpha+\xi)=R-\frac{f}{2f'}   \label{F2} \,.     
\end{align}
%
\begin{remark}
A consequence of Eq.\eqref{Riem2} is that $u^k$ is Riemann-compatible \cite{Jordan}, that is:
$$ u_i R_{jklm}u^m + u_j R_{kilm} u^m + u_k R_{ijlm} u^m =0\,. $$
It has been proven \cite{Derdz} that this property also implies that $u^k$ is Weyl compatible:
$$ u_i C_{jklm}u^m + u_j C_{kilm} u^m + u_k C_{ijlm} u^m =0\,. $$
It follows that $C_{jklm}u^m = u_k E_{jl} -u_j E_{kl}$, where $E_{kl}=C_{jklm} u^ju^m$ is the electric part
of the Weyl tensor.
\end{remark}
\section{The shear $\mbox{\large $\mathbf {\sigma_{kl}}$}$ }
\begin{lem}
Let us prove that 
\begin{align}
& \nabla^k\sigma_{kl} = u_l (\sigma_{km}\sigma^{km}) \label{divsigma}\,,\\
& \nabla_k \varphi = -\dot\varphi u_k  \,.
\end{align}
\begin{proof}
Since $\nabla^k C_{jklm}=0$, $\nabla_k(B\alpha)=(B'\nabla_k R)\alpha - B\dot\alpha u_k =
(B'\alpha^2-B\dot\alpha)u^k$ and $u^k\sigma_{jk}=0$ one evaluates:
$$B\alpha \nabla^k \sigma_{kl} = -\frac{n-2}{n-1} C_{jklm} \nabla^k (u^ju^m) =  -\frac{n-2}{n-1} u^j C_{jklm} \sigma^{km} \,.$$ 
Now, let us  use the property (see the above Remark) $u^j C_{jklm} = u_l E_{km} - u_m E_{lk} $:
$$B\alpha \nabla^k \sigma_{kl} =  -\frac{n-2}{n-1} u_l E_{km} \sigma^{km} =u_l B\alpha \sigma_{km}\sigma^{km}\,.$$ The second statement results from the identity $R_{jk}u^j =\nabla^2 u_k -\nabla_k\nabla^j u_j$, i.e.
\begin{align*}
 \xi u_k =& \nabla^l(\varphi (u_lu_k+g_{lk}+\sigma_{lk}) - (n-1)\nabla_k \varphi \\
  =& \dot\varphi u_k +(n-1) \varphi^2 u_k +\nabla_k \varphi +\nabla^l\sigma_{kl} -(n-1)\nabla_k \varphi \\
   =& [\dot\varphi  +(n-1) \varphi^2 + \sigma_{ij}\sigma^{ij}]u_k  -(n-2)\nabla_k \varphi \,.
 \end{align*}
The contraction with $u^k$ gives:
\begin{align}
\xi = (n-1)(\varphi^2+\dot\varphi ) +\sigma_{kl}\sigma^{kl}  \label{xixi}\,,
\end{align}
and the previous equation simplifies to $(n-2) (\nabla_k \varphi +u_k\dot\varphi ) =0 $.
\end{proof}
\end{lem}
It is now possible to  obtain an equation for the shear. Eq.\eqref{Riem2} simplifies with the expression \eqref{Ricci2} of the Ricci tensor:
\begin{align*}
[\nabla_j,\nabla_k]u_l = u_k \left[\frac{\xi}{n-1}g_{jl}-B\alpha \sigma_{jl}\right ]- u_j \left [\frac{\xi}{n-1}g_{kl}-B\alpha \sigma_{kl} \right ]\,.
\end{align*}
The left-hand side, with the aid of the expression for $\nabla_j u_k $, is:
$$(\dot \varphi + \varphi^2)(u_k g_{jl}-u_jg_{kl}) + \varphi (u_k \sigma_{jl}-u_j\sigma_{kl}) +\nabla_j \sigma_{kl}-
\nabla_k \sigma_{jl}\,. $$
We then obtain, with \eqref{xixi}:
\begin{align}
\nabla_j \sigma_{kl}-\nabla_k \sigma_{jl} =   (\varphi +B\alpha )(u_j\sigma_{kl}-u_k\sigma_{jl})
- \frac{\sigma_{rs}\sigma^{rs}}{n-1}(u_j g_{kl}-u_k g_{jl})\,.
\end{align}
The contraction with $u^j$ and Eq. \eqref{divsigma} gives: 
\begin{align}
\dot\sigma_{kl} + \sigma^2_{kl} + (2\varphi +B\alpha )\sigma_{kl} =
\frac{\sigma_{rs}\sigma^{rs}}{n-1} (g_{kl}+u_ku_l) \label{23}\,.
\end{align}
This equation, considered in a useful coordinate frame, will imply that $\sigma_{jk}=0$.\\

\section{The comoving frame}
Since $u^k$ is vorticity-free and acceleration-free, in the
coordinates ($t,x^1,...,x^{n-1}$) where $u_0=-1$ and $u_\mu =0$, the metric of the Lorentzian manifold has the block structure \cite{Coley} eq.2.19:
\begin{align*}
g_{ij}dx^idx^j = -dt^2 + a_{\mu\nu} (t, x) dx^\mu dx^\nu\,,
\end{align*}
where, at any time, the metric $a_{\mu\nu}$ is Riemannian.
With the formulae in Appendix, the relation $(n-1)\varphi = \nabla_k u^k$ becomes:
\begin{align}
 (n-1)\varphi =  \Gamma_{\mu 0}^\mu =\tfrac{1}{2} a^{\mu\nu}\dot a_{\mu\nu}\,. 
 \end{align}
The relation $\nabla_k u_j=\varphi (u_iu_j+g_{ij})+\sigma_{ij}$ gives $\sigma_{00}=0$, $\sigma_{0\mu}=0$ and
\begin{align}
\sigma_{\mu\nu} = \Gamma_{\mu\nu}^0 -\varphi a_{\mu\nu} = \tfrac{1}{2}\dot a_{\mu\nu} -\varphi a_{\mu\nu}\,, 
\end{align}
\begin{prop}
It is possible to show that $\sigma_{\mu\nu}=0$.
\begin{proof}
The shear is a purely spatial tensor and $\sigma_{\mu\nu} a^{\mu\nu}=0$. Let us evaluate:
\begin{align*}
\sigma_{\mu\nu}\sigma^{\mu\nu} &
=(\tfrac{1}{2}\dot a_{\mu\nu} -\varphi a_{\mu\nu} ) a^{\mu\tau}a^{\nu\sigma}  (\tfrac{1}{2} \dot a_{\tau\sigma} -\varphi a_{\tau\sigma})\\
&=\tfrac{1}{4}\dot a_{\mu\nu} a^{\mu\tau} a^{\nu\sigma}\dot a_{\tau\sigma} - \varphi \dot a_{\mu\nu}a^{\mu\nu} +\varphi^2 (n-1)\\
&=-\tfrac{1}{4}\dot a_{\mu\nu} \dot a^{\mu\tau} a^{\nu\sigma} a_{\tau \sigma} - 2\varphi^2 (n-1) +\varphi^2 (n-1)\\
&=-\tfrac{1}{4}\dot a_{\mu\nu} \dot a^{\mu\nu}   - (n-1)\varphi^2 \,.
\end{align*}
We used $0=\dot a_{\mu\nu}a^{\nu\rho} + \dot a^{\nu\rho} a_{\mu\nu} $.  \\
The equation for the shear tensor is
\begin{align*}
0=\dot\sigma_{\mu\nu} +\sigma^2_{\mu\nu} +(2\varphi +B\alpha)\sigma_{\mu\nu}-\frac{\sigma_{rs}\sigma^{rs}}{n-1}a_{\mu\nu}\,.
\end{align*}
The trace is
\begin{align*}
0=\,&\dot\sigma_{\mu\nu} a^{\mu\nu} +\sigma^2_{\mu\nu}a^{\mu\nu} +(2\varphi +B\alpha)\sigma_{\mu\nu}a^{\mu\nu}
-\sigma_{rs}\sigma^{rs}\\
=&-\sigma_{\mu\nu}\dot a^{\mu\nu} \\
=&-\tfrac{1}{2}\dot a_{\mu\nu} \dot a^{\mu\nu} +\varphi a_{\mu\nu}\dot a^{\mu\nu} \\
=&-\tfrac{1}{2}\dot a_{\mu\nu}\dot a^{\mu\nu} -\varphi \dot a_{\mu\nu} a^{\mu\nu} \\
=&-\tfrac{1}{2}\dot a_{\mu\nu}\dot a^{\mu\nu} -2(n-1)\varphi^2\\
=&\,2\sigma_{\mu\nu}\sigma^{\mu\nu}\,. \end{align*}
Since $\sigma_{\mu\nu}$ is symmetric, the eigenvalues are real. The vanishing of the sum  of squared eigenvalues means  $\sigma_{ij}=0$.
\end{proof}
\end{prop}
\begin{prop}
The property
$ a_{\mu\nu}(t,x) = a^2(t) g^*_{\mu\nu} (x)$
holds.
\begin{proof}
In the comoving frame $u_\mu=0$, then Eq.\eqref{Riem2} gives $R_{\mu\nu\rho 0}=0$. Its expression in Appendix 
shows that $\dot a_{\mu\nu}$ is the Codazzi tensor: $D_\mu \dot a_{\nu\rho} = D_\nu \dot a_{\mu\rho}$.\\
Since $\sigma_{\mu\nu}=0$,  we obtain 
$$\dot a_{\mu\nu} (t,x) = 2\varphi a_{\mu\nu}(t,x) \,.$$ 
Then $a_{\nu\rho} D_\mu \varphi = a_{\mu\rho} D_\nu \varphi $; this is true if $\partial_\nu\varphi =0 $ i.e. $\varphi $ only depends on time. 
Integration gives the warped expression $a_{\mu\nu} (t,x) =a^2(t) g^*_{\mu\nu}(x)$, where
$$\varphi (t) = \frac{\dot a}{a}\,, $$  
which is nothing else but the Hubble parameter of Friedman cosmology.
\end{proof}
\end{prop}

\section{Discussion and Conclusions}
In this paper, we discussed some properties of vacuum solutions of $f(R)$ gravity. 
Specifically, if the shear $\sigma_{jk}$ vanishes, several  formulae result simplified. In particular,  the Weyl term in the Ricci tensor \eqref{Ricci2} cancels and the Ricci tensor  becomes quasi-Einstein as in Eq. \eqref{main1}. Finally, the velocity has the simplest
expression for the gradient \eqref{nabla}.\\
Furthermore, the vanishing of the  Weyl tensor divergence implies $C_{jklm}u^m=0$ and $E_{kl}\equiv C_{jklm}u^ju^m=0$. In the 
comoving frame, $E_{kl}=0$ and $a_{\mu\nu} (t,x)= a^2(t) g^*_{\mu\nu}(x)$ give a space-submanifold that is Einstein:
$$ r_{\mu\nu} = \frac{r^*}{n-1}  g^*_{\mu\nu}\,,$$
with $r^*= r a^2$ constant and
 \begin{align}
 R=\frac{r^*}{a^2}+2\xi +\varphi^2 (n-1)(n-2)\,.
 \end{align}
The eigenvalue is $\xi = (n-1)(\varphi^2+\dot\varphi)=(n-1)\frac{\ddot a}{a}$. The parameter $\alpha $ is evaluated as $\alpha = -\dot R$.\\
The parameters $\alpha $, $R$, $\xi$ and $\varphi $ can all be expressed in terms of the constant $r^*$ and
of derivatives of the scale parameter
$a^2$. The $f(R)$ gravity enters in the relations \eqref{F1} and \eqref{F2}, that contain
$f$ and its derivatives in $R$ up to $f'''$:
\begin{align}
&(n-1)f'' \dot R \varphi = \xi f' -\frac{f}{2}  \label{FR1} \\
&f'''\dot R^2 + f''\ddot R +\xi f' = \frac{1}{n-1}(R f'-\frac{f}{2} ) \label{FR2}
\end{align}
The first equation reproduces Eq.10 in \cite{CMM2018} in vacuo ($\mu =0$); the second equation is Eq.9 (with $p=0$) plus
$\frac{n-2}{n-1}$ times Eq.10 in  \cite{CMM2018}.

According to these considerations, if the divergence of the Weyl tensor is zero, the vacuum solutions of $f(R)$ gravity are  Generalized Friedman-Robertson-Walker space-times which, in  4-dimensions,  reduce to the standard Friedman-Robertson-Walker solutions. This result holds, in particular, for $f''\neq 0$  and then it generalizes the fact that  $f(R)$ gravity in vacuum can be reduced to General Relativity plus a cosmological constant as often stated. In the present perspective, the properties of the Weyl tensor and the geodesic structure determine the solutions.\\
In 4-dimensions, where the FRW metric is obtained, the difference with ordinary gravity is in the equations for the evolution of the scale function: the Friedmann equations for $f(R)=R$ and equations \eqref{FR1}, \eqref{FR2} for $f(R)$ with null divergence of the Weyl tensor. Therefore, the effect of $f(R)$ with Weyl constraint is to dress the original Friedmann equations with new geometric effects
that yield the same form of metric and manifest in a different scale function.

There are cosmological spaces that elude this analysis, because of physical processes \rem{working in the realistic universe} that do not allow the stringent requirement on the Weyl tensor to hold. For example, this is  the case of
de Sitter solutions derived in $f(R)$ gravity motivated by the inflation and the dark energy issues. In Ref.\cite{Cognola}, the one-loop quantization approach is developed for a family of $f (R)$ gravity models and de Sitter universes are investigated, extending a similar earlier program for Einstein gravity. The authors adopt a generalized zeta regularization, and  the one-loop effective action is  obtained off-shell. In this framework,  the (de)stabilization of the de Sitter background is obtained by quantum effects. In such a context, the requirements of Theorem 1 are violated by quantum 
effects. In fact, they consider small fluctuations around a (de Sitter) maximally symmetric space with Riemann
tensor
$$ R^{(0)}_{jklm}=\frac{R^{(0)}}{12}(g_{jl}^{(0)}g_{km}^{(0)}-g_{jm}^{(0)}g_{kl}^{(0)}) $$
and metric $g_{ij}^{(0)}$. In this metric the Weyl tensor is zero. However, in a perturbed metric $g_{ij} =
g_{ij}^{(0)} + h_{ij}$ the shape of the Ricci tensor (eq. 2.11 in \cite{Ginsparg}) in general prevents the validity of
$\nabla_m C_{jkl}{}^m=0$. \\
It is important to stress that one-loop effective actions are useful also for studying  black hole nucleation rates  and for providing reliable mechanisms capable of  solving the cosmological constant problem.

In conclusion, the mathematical results reported here highlight the decisive effect of the Weyl condition in 
restricting the form of the vacuum solutions, and this should be of guidance in the study of realistic situations where physical processes are acting in the observable universe.

\begin{acknowledgements}   
SC acknowledges {the} Istituto Nazionale di Fisica Nucleare (INFN), sezione di Napoli, {\it iniziative specifiche} MOONLIGHT2 and  QGSKY. 
\end{acknowledgements}

\section*{APPENDIX}
\label{appendix}
Given the space-time metric 
$$ ds^2 =-dt^2 + a_{\mu\nu} (t,x) dx^\mu dx^\nu $$
let $\gamma_{\mu\nu}^\rho $, $r_{\mu\nu\rho}{}^\sigma $, $r_{\mu\nu}$ and $r$ be the Christoffel symbols, the Riemann tensor,
the Ricci tensor and the Ricci scalar of the space-submanifold at fixed $t$, and let $D_\mu $ be the covariant derivative with
symbols $\gamma_{\mu\nu}^\rho $, and define $a^{\mu\nu}a_{\mu\rho}=\delta^\mu{}_\rho $. 
The related space-time quantities are:
\begin{align*}
&\bullet \text{\sf Christoffel symbols:}\;\Gamma_{ij}^k = \Gamma_{ji}^k=\tfrac{1}{2} g^{km}(\partial_i g_{jm} + \partial_j g_{im} -\partial_m g_{ij})\\
&\Gamma_{00}^0 =0,\quad \Gamma_{\mu 0}^0 =0, \quad \Gamma_{00}^\mu =0,\\
&\Gamma_{\mu\nu}^0 = \tfrac{1}{2}\dot a_{\mu\nu}, \quad \Gamma_{\mu 0}^\nu = \tfrac{1}{2} a^{\nu\rho}\dot a_{\mu\rho},
\quad \Gamma_{\mu\nu}^\rho =\gamma_{\mu\nu}^\rho\\
\\
&\bullet \text{\sf Riemann tensor:}\;R_{jkl}{}^m =\partial_k \Gamma_{jl}^m -\partial_j \Gamma_{kl}^m + \Gamma_{jl}^i \Gamma_{ik}^m - \Gamma_{kl}^i \Gamma_{ij}^m\\
&R_{\mu\nu\rho}{}^\sigma =r_{\mu\nu\rho}{}^\sigma
+ \tfrac{1}{4} a^{\sigma\lambda}(\dot a_{\mu \rho}\dot a_{\lambda\nu} -  \dot a_{\nu\rho} \dot a_{\lambda\mu}) \\
&R_{\mu\nu\rho}{}^0 =\tfrac{1}{2}(\partial_\nu \dot a_{\mu \rho} -\partial_\mu \dot a_{\nu \rho}) +\tfrac{1}{2} (\gamma_{\mu \rho}^\sigma \dot a_{\nu\sigma} - \gamma_{\nu \rho}^\sigma \dot a_{\sigma\mu})  = \tfrac{1}{2}(D_\nu \dot a_{\mu\rho} -
D_\mu \dot a_{\nu\rho})\\
&R_{0\mu0}{}^\nu = -\tfrac{1}{2} a^{\nu\rho} \ddot a_{\mu\rho}  -\tfrac{1}{4} \dot a^{\nu\rho}\dot a_{\mu\rho} \\
&R_{\mu 0 \nu}{}^0 = \tfrac{1}{2} \ddot a_{\mu\nu} -\tfrac{1}{4} a^{\rho\sigma}\dot a_{\nu\sigma} \dot a_{\mu\rho }\\
\\
&\bullet \text{\sf Ricci tensor:}\; R_{ij}=R_{ikj}{}^k\\
&R_{00} = -\tfrac{1}{2} a^{\mu\nu} \ddot a_{\mu\nu}-\tfrac{1}{4} \dot a^{\mu\nu}  \dot a_{\mu\nu} \\
&R_{\mu\nu} =r_{\mu\nu} - \tfrac{1}{2}\dot a_{\mu\rho} a^{\rho\sigma}
\dot a_{\nu\sigma} +\tfrac{1}{2} \ddot a_{\mu\nu} +\tfrac{1}{4} \dot a_{\mu\nu} (a^{\rho\sigma}\dot a_{\rho\sigma} )\\
\\
&\bullet \text{\sf Curvature scalar}\; R= -R_{00} +a^{\mu\nu}R_{\mu\nu}\\
&R =
r +a^{\mu\nu}\ddot a_{\mu\nu} + \tfrac{3}{4} \dot a^{\mu\nu}  \dot a_{\mu\nu} + \tfrac{1}{4} (a^{\mu\nu} \dot a_{\mu\nu})^2\\
\\
&\bullet \text{\sf Electric tensor}\; E_{\mu\nu} =C_{0\mu\nu 0}\\
&(n-2)E_{\mu\nu} = -r_{\mu\nu} +\tfrac{1}{2}(n-3)\ddot a_{\mu\nu}  -\tfrac{1}{4}(n-4)a^{\rho\sigma}\dot a_{\rho\mu}\dot a_{\sigma\nu} -\tfrac{1}{4}\dot a_{\mu\nu}(a^{\rho\sigma} \dot a_{\rho\sigma}) + a_{\mu\nu}(R_{00}+\tfrac{R}{n-1})
\end{align*}

\vfill
\end{document}